\documentclass[runningheads]{llncs}
\usepackage[T1]{fontenc}
\usepackage{graphicx}
\usepackage{cite}
\usepackage{amsmath,amssymb,amsfonts}
\usepackage{algorithmic}
\usepackage{graphicx}
\usepackage{textcomp}
\usepackage{xcolor}
\usepackage{booktabs}
\usepackage[hidelinks]{hyperref}
\usepackage{multirow}
\usepackage{url}

\usepackage{orcidlink} 

\begin{document}
\title{Feasibility Study for Supporting Static Malware Analysis Using LLM}
\author{
  Shota Fujii\inst{1}\orcidlink{0000-0002-1250-7488}
  \and
  Rei Yamagishi\inst{1}\orcidlink{0000-0002-0131-5704}
}
\authorrunning{S. Fujii et al.}
\institute{
Hitachi, Ltd., Research and Development Group, Kanagawa, Japan\\
\email{\{shota.fujii.xh,rei.yamagishi.ss\}@hitachi.com}
}
\maketitle              
\begin{abstract}
Large language models (LLMs) are becoming more advanced and widespread and have shown their applicability to various domains, including cybersecurity. 
Static malware analysis is one of the most important tasks in cybersecurity; however, it is time-consuming and requires a high level of expertise. 
Therefore, we conducted a demonstration experiment focusing on whether an LLM can be used to support static analysis. 
First, we evaluated the ability of the LLM to explain malware functionality. The results showed that the LLM can generate descriptions that cover functions with an accuracy of up to 90.9\%. 
In addition, we asked six static analysts to perform a pseudo static analysis task using LLM explanations to verify that the LLM can be used in practice. 
Through subsequent questionnaires and interviews with the participants, we also demonstrated the practical applicability of LLMs. 
Lastly, we summarized the problems and required functions when using an LLM as static analysis support, as well as recommendations for future research opportunities.

\keywords{Malware \and Static analysis \and LLM \and Usable security.}
\end{abstract}

\section{Introduction}
Malware used in cyberattacks has become more sophisticated and widespread over the years. 
Under these circumstances, it has become even more important to analyze malware and take countermeasures as quickly as possible. 
In some cases, reverse engineering is used to analyze suspicious software or malware that may be related to one's own organization, and the behavior of the malware is analyzed in detail in what is called static analysis. 
Static analysis is not easily affected by the evasion of the analysis environment and plays an important role in understanding malware behavior. 
However, static analysis is known to be time-consuming as it is often performed manually \cite{Yakdan2016,Mantovani2023}. 
Although there are methods for assigning types and variable names to decompiled results based on the content of the function \cite{Chen2022}, most of the analysis is still done manually. 
Tools for supporting and automating analysis have been developed but are not widely used in practice due to their low usability \cite{Mattei2022}.

In this paper, we discuss methods for improving the efficiency of static analysis. 
Specifically, we investigate the possibility of using ChatGPT \cite{ChatGPT} (GPT-4), one of the recently developed large language models (LLMs), to generate explanatory text about malware functions to support static analysis. 
For this verification, the results of decompiling/disassembling the malware for each function were entered into ChatGPT to generate a description, and then the coverage for the malware functions described in the published articles was evaluated. 
We then tried different types of prompts (commands to LLM) and also verified which type of prompts were desirable for generating malware descriptions.

Although it would be useful to be able to describe malware in a form similar to that of an article, it is not equivalent to supporting all malware static analysis. 
Therefore, we asked static analysts to solve a task simulating malware static analysis using ChatGPT explanations in addition to the decompiled/disassembled results and evaluated its practicality through questionnaires and interviews. 
We also tried to derive the problems and system requirements that can be expected in practical use, towards real-world applications in the future.

The main contributions of this research are as follows:

\begin{itemize}
  \item 
    To support the static analysis of malware, we designed several prompts, generated malware function explanations using an LLM, and quantitatively evaluated the explanations using functional coverage and agreement with the analysis article. 
    Specifically, we showed that the LLM can generate explanations that cover functions with up to 90.9\% accuracy, and through these results we demonstrated the potential of LLM usage.
  \item 
    Six static analysts were asked to perform a simulated static analysis task using LLM descriptions. 
    The subsequent questionnaires and interviews also demonstrated the potential of LLM use from a practical perspective. 
  \item 
    Based on the analysis of the questionnaire and interview results, we have developed a proposal for future research in the area of LLMs, along with problems, and required functions and user interfaces when using LLM-generated results as static analysis support.
\end{itemize}

\begin{figure}[t]
 \begin{center}
 \includegraphics[width=1.0\columnwidth]{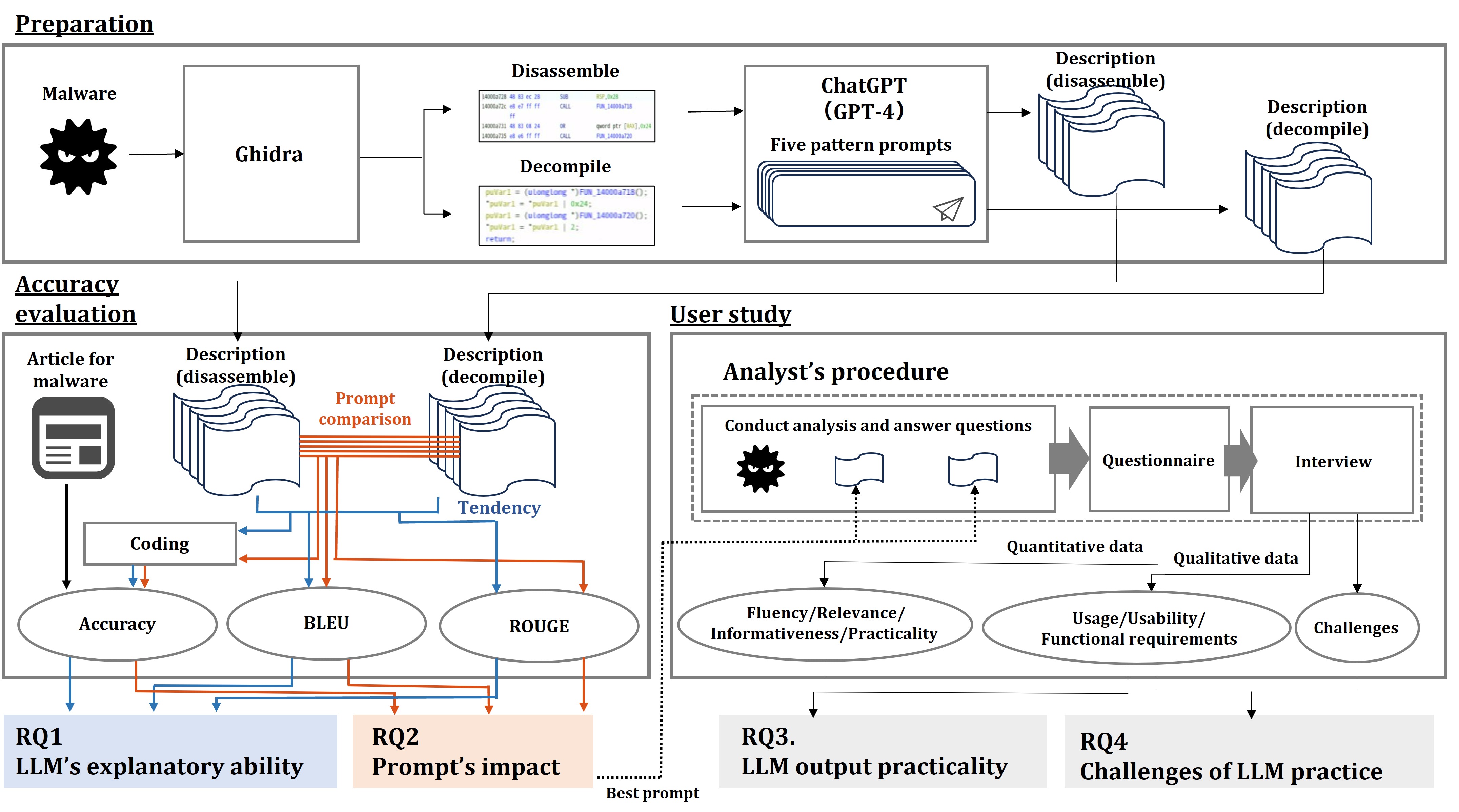}
 \caption{Overview of our study.}
 \label{fig:overview}
 \end{center}
\end{figure}

\section{Background}

\subsection{Static Malware Analysis}

Malware analysis methods include surface analysis, which analyzes malware based on superficial information; dynamic analysis, which analyzes malware based on its behavior during actual operation; and static analysis, which analyzes the results of decompiling/disassembling malware. 
Static analysis is particularly important for understanding malware behavior because it is less susceptible to circumvention of the analysis environment.

However, static analysis is often performed manually and is known to be time-consuming \cite{Yakdan2016,Mantovani2023}. 
Although there are methods for assigning types and variable names to decompiled results based on the content of the function \cite{Chen2022}, most of the analysis is still done manually. 
Although support tools and automation tools are available, they are not widely used in practice due to their lack of usability \cite{Mattei2022}.

\subsection{Large Language Models}
LLMs have shown their potential to support a variety of tasks. 
In particular, LLMs have demonstrated high zero-shot performance in natural language processing tasks such as translation and summarization.

LLMs are also being applied to various domains including cybersecurity, for example, to fix vulnerable codes \cite{Pearce2023} and find the root cause of incidents in cloud environments \cite{Ahmad2023}.

\subsection{Research Challenges}
As mentioned above, static analysis of malware entails extensive human intervention and high implementation costs. 
One way to alleviate these problems is to use LLMs to support static analysis tasks. 
To advance this study, we formulated and verified the following four research questions (RQs).

\textbf{RQ1. Can LLMs generate explanations that contribute to static analysis?} 
Mantovani et al. \cite{Mantovani2023} state that in static analysis (reverse engineering), it is important to clarify the function of the analysis target. 
Therefore, we evaluated the coverage of LLM-generated malware function explanations to the functions described in the malware analysis articles. 
In addition, we compared the analysis articles and the text generated by LLM and evaluated the degree of agreement using various evaluation indices.

\textbf{RQ2. To which extent prompts do affect explanatory ability of LLMs?} 
In general, the form and wording of prompts given to LLMs are known to affect task performance \cite{jiang2020}. 
This has also been found to be true in the domain of cybersecurity, e.g., when fixing vulnerable code \cite{Pearce2023}. 
Therefore, we designed several prompts and investigated the differences in their accuracy, as well as the desirable prompts for static analysis support.

\textbf{RQ3. How useful are LLM outputs for analysts?}
It can be inferred that an LLM's ability to provide explanations that contribute to static analysis is not necessarily equivalent to its ability to assist real analysts. 
Therefore, we asked static analysts to work on a task simulating static analysis of malware using LLM explanations and evaluated the practicality and usefulness of LLM output through questionnaires and interviews.

\textbf{RQ4. What are the challenges with the practical application of LLM support?} 
When using LLM explanations to support static malware analysis, it can be inferred that there are issues that may hinder practical use and issues that can be resolved to improve usability.
Therefore, we interviewed analysts to find out what issues they might face in the future.

\begin{table}[t]
  \caption{Malware for evaluation.}
  \begin{center}
  \scalebox{0.70}{
    \begin{tabular}{lllll}
      \hline
      \hline
      Family & Type & Hash Value (SHA-256) & Extension & Functions \\
      \hline
      Babuk & Ransomware & 3ab167a82c817cbcc4707a18fcb86610090b8a76fe184ee1e8073db152ecd45e & EXE & 62/107 \\
      \hline
    \end{tabular}
  }
  \label{tab:sample_list}
  \end{center}
\end{table}
 
\begin{table}[t]
  \caption{List of prompts for evaluation.}
  \begin{center}
  \scalebox{0.70}{
    \begin{tabular}{rl}
      \hline
      \hline
      PID & Prompt \\
      \hline
      1 & Please explain the following function: \textbf{[code]} \\
      \hline
      2 & Act as a malware analyst and explain the following function: \textbf{[code]} \\
      \hline
      3 & Please summarize the following function: \textbf{[code]} \\
      \hline
      4 & You are a malware analyst. Please explain the following function focusing on the suspicious parts: \textbf{[code]} \\
      \hline
      \multirow{2}{*}{5} & The following is the [decompiled \textbar disassembled] function identified by Ghidra from malware. \\
       & Please explain this function as you would in a malware analysis article: \textbf{[code]} \\
      \hline
    \end{tabular}
  }
  \label{tab:prompt_list}
  \end{center}
\end{table}

\section{Methodology}
\label{sec:design}

\subsection{Overview}

This section describes the research methodology used to verify the aforementioned RQs.

Figure \ref{fig:overview} shows an overview of this study. 
First, we select malware to evaluate for verification. 
Next, we evaluate the accuracy to verify RQ1 and RQ2. 
To validate RQ3 and RQ4, we requested the cooperation of static analysts and conducted a user study through questionnaires and interviews. 
The details of each of these are described in the following sections.

In designing the methodology, we developed the procedure with reference to \cite{Mantovani2023} for items related to static analysis, \cite{Ahmad2023,Pearce2023} for items related to LLMs in the cybersecurity field, and \cite{Roth2021,Wermke2022} for items related to other user studies.

\subsection{Malware Selection}

The malware used in the evaluation was selected based on the following criteria.

\begin{enumerate}
 \item[(1)] The analytical article for the malware must be publicly available.
 \item[(2)] The hash value of the malware must be included in the analysis article.
 \item[(3)] It must be possible to decompile/disassemble the malware.
\end{enumerate}

First, we obtained articles from various security vendors during the period from November 2022 to May 2023. 
After that, to extract articles that satisfied condition (1) above, we extracted 33 articles whose article category was \textit{malware} or whose headlines contained words related to analysis, such as \textit{analysis}. 
Finally, after reviewing the article content and malware details, one malware that also met conditions (2) and (3) was extracted and selected for this evaluation. 
Note that the selected malware is called \textit{Babuk} and its analysis article was written by Cisco Talos \cite{babuk_article}.
The details of the selected malware are shown in Table \ref{tab:sample_list}.

For evaluation, we decompiled/disassembled the selected malware using Ghidra \cite{Ghidra} 10.3 and used each result as input to the LLM to generate an explanation using the prompts described below. 
However, when the entire malware was targeted, the length of the decompile/disassemble results greatly exceeded the input limit of the LLM. 
Thus, this time we generated explanations for each function extracted by Ghidra by splitting them into separate sections. 
We excluded library functions such as \verb|strcmp| because their behavior is obvious and not subject to analysis, and focused only on the malware's own functions. 
Since Ghidra can detect library functions using hash values, library functions have been excluded by this function.
The results are listed in the \textit{Function} column of Table \ref{tab:sample_list}. 
Specifically, this column shows 62/107, meaning that the malware in the table has a total of 107 functions, 62 of which are malware-specific functions that are the target of the analysis.

\subsection{Accuracy Evaluation}
\label{subsec:eval_accuracy}
To validate RQ1, we evaluated the extent to which the explanatory text generated by the LLM covered the functions of the malware being evaluated. 
We also evaluated the extent to which the explanatory text matched the explanatory article for the same malware.

First, to evaluate the coverage of the functions, we read the article describing the malware and the LLM output and extracted the sections describing the functions of the malware. 
In this process, coding was performed to formulate the functions of the malware. 
Specifically, following the recommendations in \cite{Ortloff2023}, a lead author first created a codebook, and then the two authors read the articles describing the malware and the LLM output, and extracted the functions described in each article. 
After completing each task, the two authors discussed and agreed on the results, which were used as the correct data. 
It was determined that 11 functions of the malware were described in the explanatory article. 
We quantitatively evaluated the extent to which the output of the LLM covered 11 of these functions using the aforementioned coding results. 
Note that the codebook created is shown in Table \ref{tab:malware_functions} in Section \ref{sec:appendix_a} of the appendix.

To measure the degree of agreement between the results generated by the LLM and the malware analysis article, we performed an accuracy evaluation using BLEU \cite{BLEU} (BLEU-4) and ROUGE \cite{ROUGE} (ROUGE-1, Recall), which are widely used in the field of text generation. 
In this evaluation, the ground truth was the sentences contained in the malware analysis article. 
BLEU is a widely used metric for evaluating models such as machine translation. It evaluates how many of the N-grams in the generated text are included in the correct text. 
ROUGE is a widely used metric for evaluating models such as summary generation. It evaluates how many of the N-grams of the correct summary is included in the generated summary. 
Using each metric, we evaluated the degree of agreement between the results generated by the LLM and the text in the commentary article as the correct answer. 
This evaluation was conducted using English explanatory articles and English generated sentences.

To further validate RQ2, multiple prompts were designed, and the aforementioned evaluation was performed for each prompt. 
In this experiment, a policy was formulated with reference to \cite{coyne2023,Pearce2023}, which focused on differences in accuracy across prompts, and five prompts were prepared. 
Specifically, a simple prompt was given first. 
Next, we set up a prompt that specified the user's role as a malware analyst and asked them to explain the function of malware. 
We also set up prompts that provided an explanation, a summary, specific points to focus on, and background information.

The actual prompts are shown in Table \ref{tab:prompt_list}. 
The explanatory text is generated by assigning the decompile or disassemble result to the \textbf{[code]} part at the end of each prompt. 
We then evaluated the accuracy of a total of 10 patterns in which each prompt was used to generate explanatory text for each of the decompile/disassemble results.

\subsection{User Study}

\subsubsection{Recruitment}

To validate RQ3 and RQ4, we conducted a user study in addition to the accuracy evaluation mentioned above. 
Prior to the user study, we prepared a recruitment letter and poster, following \cite{Roth2021}, and recruited participants through the analyst community within group companies. 
Specifically, the recruitment included the purpose of the study, the requested items, the estimated time required, and the target participants. 
As a result, the user study was conducted with the cooperation of six analysts from four organizations, as shown in Table \ref{tab:p_list}.

\begin{table}[t]
  \caption{List of participants}
  \begin{center}
  \scalebox{0.70}{
    \begin{tabular}{lllll}
      \hline
      \hline
      Participant ID & Organization ID & Analysis experience & Security experience & Frequency of static analysis \\
      \hline
      P1 & A & 2--3 years & 5--7 years & Sometimes \\
      P2 & B & 10--15 years & 10--15 years & Frequently \\
      P3 & B & 7--10 years & 7--10 years & Sometimes \\
      P4 & A & 2--3 years & Over 15 years & Only competition \\
      P5 & C & 4--5 years & 5--7 years & Sometimes \\
      P6 & D & 5--7 years & Over 15 years & Sometimes \\
      \hline
    \end{tabular}
  }
  \label{tab:p_list}
  \end{center}
\end{table}

\begin{figure}[t]
  \begin{center}
  \includegraphics[width=1.0\columnwidth]{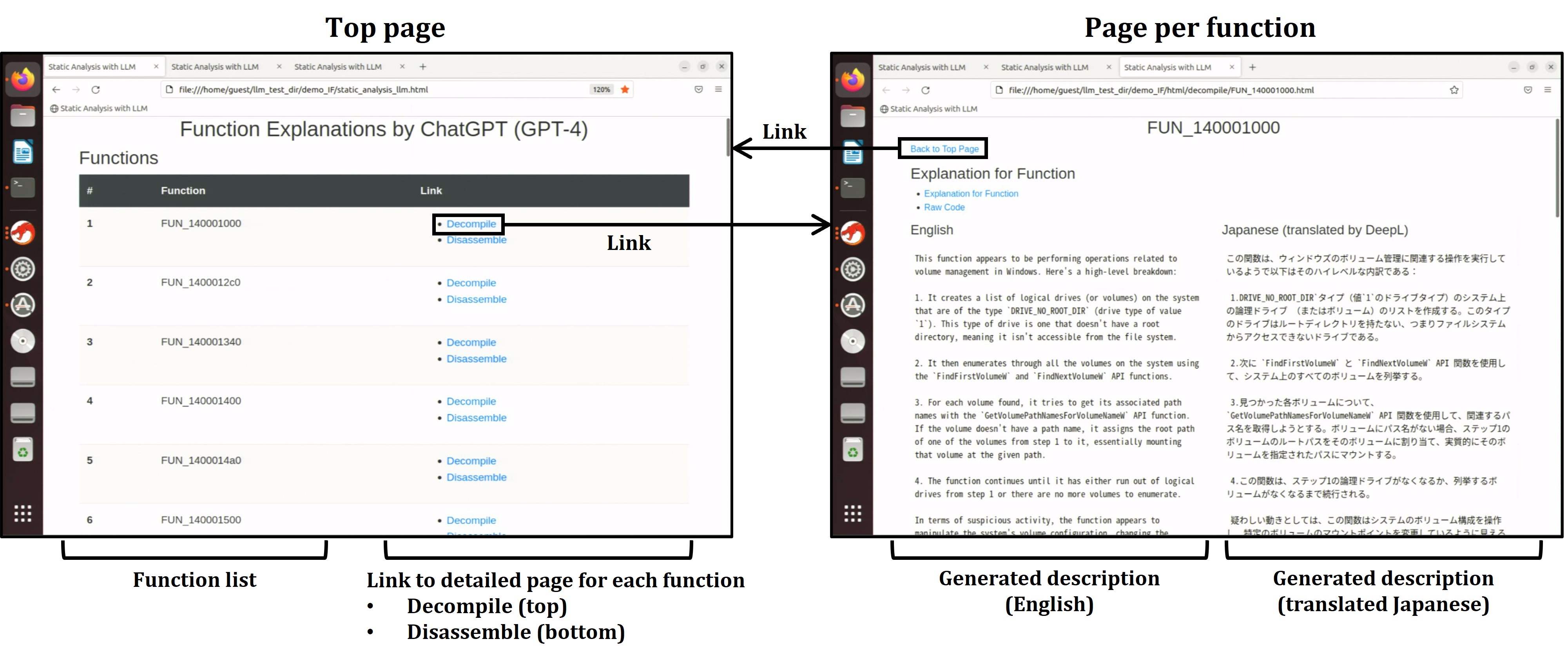}
  \caption{User interface displaying a description of each function of the malware generated by the LLM.}
  \label{fig:experimental_environment}
  \end{center}
\end{figure}

\subsubsection{Procedure}
The procedure for conducting user study consisted of the following steps: preliminary explanation, analysis, post-test questionnaire, and interview.

\textbf{1. Preliminary explanation.} 
The purpose and details of the study were explained to the participants in advance. 
We then distributed materials on how to use Ghidra and how to connect it to the experimental environment.

\textbf{2. Analysis.} 
For the analysis, a PC with the necessary software such as Ghidra installed in an isolated virtual environment was provided to the participants as a unified analysis environment. 
In addition, we provided a simple system that allows users to view the list of functions and LLM explanations for the disassemble/decompile results of each function via a web browser. 
The interface of the system is shown in Fig. \ref{fig:experimental_environment}. 
Participants had the choice of connecting to the analysis environment remotely or locally with a virtual machine image.

\begin{table}[t]
  \caption{Accuracy evaluation results per prompt}
  \begin{center}
  \scalebox{0.70}{
    \begin{tabular}{rrlrrrrrlrrr}
      \hline
      \hline
      \multirow{2}{*}{PID} & \multicolumn{5}{c}{Decompile} & & \multicolumn{5}{c}{Disassemble} \\
      \cline{2-6}
      \cline{8-12}
      & \multicolumn{2}{c}{Accuracy} & BLEU & ROUGE & Average words per function & & \multicolumn{2}{c}{Accuracy} & BLEU & ROUGE & Average words per function\\
      \hline
      1 & 9/11 & (81.8\%) & 0.31 & 57.9 & 1928.8 & & 1/11 & (9.1\%) & 0.32 & 50.3 & 1901.4\\
      2 & 10/11 & (90.9\%) & 0.43 & 61.9 & 1955.2 & & 2/11 & (18.2\%) & 0.19 & 57.4 & 2103.0\\
      3 & 9/11 & (81.8\%) & 0.46 & 55.8 & 1343.8 & & 1/11 & (9.1\%) & 0.30 & 48.7 & 1521.2\\
      4 & 10/11 & (90.9\%) & 0.45 & 59.9 & 2059.9 & & 3/11 & (27.3\%) & 0.42 & 59.4 & 2004.6\\
      5 & 10/11 & (90.9\%) & 0.40 & 64.0 & 2077.1 & & 1/11 & (9.1\%) & 0.39 & 57.7 & 2319.4\\
      \hline
    \end{tabular}
  }
  \label{tab:accuracy}
  \end{center}
\end{table}
 
Participants were asked to answer questions that solved a quasi-static analysis task in the environment described above, using LLM explanations. 
The questions are as follows (options are omitted in this paper).

\begin{enumerate}
 \item[Q1.] Please select this type of malware from the following candidates. (Answer: Ransomware)
 \item[Q2.] Please give the name of the function that waits for commands from the attacker's server and performs malicious actions on the victim's computer according to the content of the commands. (Answer: None)
 \item[Q3.] Please give the name of the function that will delete the shadow volume copy of the victim's computer. (Answer: FUN\_140001340)
 \item[Q4.] FUN\_14001000 scans the volumes on the computer and maps unmounted volumes to empty drives. Please select all of the following drives that are not specified as the target drive for the mapping. (Answer: All of the above drives are candidates for the mapping destination)
\end{enumerate}

The answer to Q1 is based on the classification of malware in \cite{meir2019}. 
The answers to Q2 through Q4 are designed so that the answers to the other questions cannot be uniquely identified in order to suppress the learning effect \cite{van2019} caused by the other questions. 
Note that this task is designed so that the estimated response time would be about one hour for the questions to be answered by static analysis of the malware, following \cite{Mantovani2023}.

Since it is known that English has a significant accuracy advantage for LLMs that can handle multiple languages, including ChatGPT \cite{etxaniz2023multilingual}, English was also used in this experiment. 
As all participants were native Japanese speaking analysts, the LLM output was distributed together with a Japanese translation using DeepL \cite{DeepL}.

\textbf{3. Post-test questionnaire.} 
After the analysis, a post-analysis questionnaire was distributed using Microsoft Forms \cite{Forms}. 
The questionnaire items were as follows.

\begin{itemize}
 \item History of working in the security industry.
 \item History and frequency of conducting malware analysis.
 \item Time spent on analysis.
 \item Prior knowledge of the malware.
 \item LLM results: fluency, relevance, informativeness, and practicality (1-4 Likert scale).
 \item Differences between the environment in which the analysis is usually performed and the environment in which it is distributed.
 \item Any additional comments.
\end{itemize}

First, to understand the demographics of the participants, we asked about their history of working in security, and history and frequency of conducting malware analysis. 
In addition, questions were included to determine if they were familiar with the malware prior to the analysis and any differences from their usual analysis environment that might affect the results. 
Furthermore, a Likert scale was included to verify the usefulness of the LLM outputs, with 4 being the highest and 1 being the lowest. 
Fluency, relevance, and informativeness were set up in the form of extracting the evaluation indicators \cite{van2019} in text generation that can be used in this task. 
Since one of the objectives of this user study is to verify whether the LLM outputs can be utilized in practice, a practicality perspective was added to the questions. 
For the Likert scale questions, the criteria for each value were given.

Finally, an open-ended field for additional comments was included. 
If an item was mentioned in this section, it was explored in depth later in the interview, along with the answers to the questions above.

\textbf{4. Interview.} 
A semi-structured interview was conducted with each participant and one lead author. 
The interview was held online and recorded using Microsoft Teams \cite{Teams}. 
Following the example in \cite{Wermke2022}, we prepared an interview script in advance which included the following main points and questions.

\begin{itemize}
 \item Reminder of the purpose of this interview.
 \item Reminder that the interview is being recorded.
 \item Confirmation and detailed review of interview content (e.g., reasons for each answer).
 \item Challenges of using LLMs.
 \item How LLMs will be used in this analysis.
 \item Usability of the system for viewing LLM results and functional requirements of the system.
 \item Any other topics. 
\end{itemize}

A codebook was then created by the lead author, and the interview recordings were reviewed and coded by both authors. 
After each task was completed, the authors discussed and refined the results into a data set. 
Finally, this data set and the aforementioned questionnaire were analyzed to verify the possibility of using LLMs in practice for RQ3 and RQ4.

\section{Analysis Results}

\subsection{RQ1/RQ2: LLM's Explanatory Ability and Prompt's Impact}
\label{subsec:rq12}

Table \ref{tab:accuracy} shows the results of the accuracy evaluation of the function descriptions output by the LLM for the malware under evaluation, using the function recognition accuracy, BLEU, and ROUGE. 
This evaluation was performed for each prompt listed in Table \ref{tab:prompt_list}, and the results shown in each row of Table \ref{tab:accuracy} are for the prompt corresponding to the PID in Table \ref{tab:prompt_list}.

For all evaluation metrics, the accuracy tends to be higher when decompiled results are used as input to the LLM than when disassembled results are used as input. 
For PID2, PID4 and PID5, which yielded the most accurate decompilation results, about 90\% of the malware functions can be explained.

In summary, these results suggest that the LLM is capable of generating explanations that support static analysis for this malware (RQ1).

The low accuracy of the disassembled results as input may be due to the fact that the disassembled results include heap space allocation and other items not directly related to the behavior of the function.

Of the disassemblers, PID4 had the highest accuracy. PID2, PID4, and PID5 were equally accurate for decompilation, but PID4 was most likely to describe malware. 
For example, the text file generated by the ransomware that was the target of this evaluation mentioned the possibility that the text contained a ransom note and the enumeration of logical drives in preparation for encryption, all of which were answered correctly. 
This suggests that the instruction to focus on and explain suspicious points, which is only included in PID4, had a positive effect. PID2, PID4 and PID5, which were highly accurate, all contained instructions to act as a malware analyst. 
These results are consistent with existing research showing that role-playing instructions improve LLMs' reasoning skills \cite{Kong2023}.

From these results, we concluded that prompts affect the explanatory ability of LLMs and that accuracy can be improved by having the LLM act as a malware analyst and giving instructions to focus on suspicious parts (RQ2).

Considering the highest function recognition accuracy and the significance of the explanatory text described above, the LLM output provided to the participant when requesting analysis was the one generated by PID4.

\subsection{RQ3: LLM Output Practicality}
\label{subsec:rq3}

Table \ref{tab:anketo} shows the results of the questionnaires and interviews. 
As shown in the table, fluency, relevance, informativeness, and practicality were all high.

All participants stated that fluency was not a problem. Specifically, they stated:
\begin{quote}
P1, P3, P4, P5, P6
\textit{``It's not weird at all.''}

P5, P6
\textit{``It was so easy to read that just reading the output was enough to understand the process.''}
\end{quote}

Relevance was also generally rated highly, e.g., \textit{``The output matched the processing content of the function''} (P6). 
Some participants suggested that presenting information related to the malware may have contributed to its usefulness, e.g., \textit{``The parameters of the WindowsAPI call were explained clearly.''} (P5). 
However, some participants noted that when the prompts instructed the user to act as a malware analyst, \textit{``the explanations forcefully linked unrelated items to malware behavior, which was confusing.''} (P2). 
P2 also stated that it would have been easier to understand the generated prompt if ChatGPT had been instructed to simply explain the function. 
This suggests that, from a practical standpoint, it would be desirable for users to be able to change the prompts to suit their needs.

All of participants stated that they were satisfied with informativeness; however, some participants had the following concerns:

\begin{quote}
P2 
\textit{``Some explanations for long functions were not informative enough.''}

P1, P2, P3, P4
\textit{``Non-important information was sometimes included in redundant form.''}
\end{quote}

This suggests that it is desirable to control the amount of information depending on the length and importance of the function, rather than generate explanations uniformly for all functions. 
As shown in Table \ref{tab:accuracy}, the output length can be controlled to some extent through the prompt; this is one of the issues that should be investigated further in future research.

Practicality also received a relatively high score, with an average of 3.17, and most of the comments were positive, as follows.
\begin{quote}
P5, P6
\textit{``The explanation is good and practical.''}

P4
\textit{``Sufficient as a support tool.''}
\end{quote}
However, some participants also expressed that the LLM output insufficient for the static analysis, suggesting that it is realistic to use the LLM as an adjunct to existing static analysis rather than relying only on the LLM output at this time. The participant stated: 
\begin{quote}
P2, P4
\textit{``It was necessary to check the results of decompile and disassemble together, and although the LLM output is useful, it was difficult to rely on it alone to analyze the results.''}
\end{quote}

In summary, despite some challenges, the results suggest that LLM's output explaining malware is useful to the analyst (RQ3).

\begin{table}[t]
  \caption{Results of questionnaire on LLM output}
  \begin{center}
  \scalebox{0.70}{
    \begin{tabular}{lrrrr}
      \hline
      \hline
      & Fluency & Relevance & Informativeness & Practicality \\
      \hline
      Mean & 3.83 & 3.5 & 3.17 & 3.17\\
      Median & 4 & 4 & 3 & 3\\
      \hline
    \end{tabular}
  }
  \label{tab:anketo}
  \end{center}
\end{table}

\subsection{RQ4: Challenges of LLM Application}
\label{subsec:rq4}

Through questionnaires and interviews, we were able to identify 18 issues. 
Due to space limitations, here we discuss a selection of the more important ones.
Note that all of issues are listed on Table \ref{tab:codebook_interview} in Section \ref{sec:codebook_interview} of the appendix.

First, most of the participants (P2, P3, P4, P5, and P6) raised concerns about confidentiality when using ChatGPT to send information outside the organization to receive output: \textit{``There is a possibility that information hard-coded in the malware (organization name, DNS, etc. of the target organization, and authentication information) could be leaked''}. 
In addition, P1 stated that \textit{``There may be contractual issues such as not using external services when using the system in practice''}. 
This issue needs to be addressed as a matter of priority, especially if it is used in practice.

Some participants mentioned possible output interference at the LLM, as follows:

\begin{quote}
P1, P3, P4, P6
\textit{``Possible obstructions, such as obfuscation, could reduce the accuracy of the explanation.''}

P4, P6
\textit{``There is a possibility of potential interference from junk code.''}
\end{quote}

Such obfuscation, which reduces the accuracy of explanations and causes redundant or useless explanations to be generated by junk code (useless code or false instructions), is an obstacle to the efficiency of static analysis that this research aims to achieve, and is an issue that needs to be addressed in future work.

In addition, we received feedback that it would be better to have other outputs in addition to an explanation of each function.
Specifically, we received feedback that a general description of the malware as a whole could help the analyst get an overview of the malware first, thus improving the efficiency of the analysis process.
Mantovani et al. \cite{Mantovani2023} also found that analysts first get a general overview before proceeding to a detailed analysis. Providing support that is tailored to the analyst's behavior and the analysis phase can further support and improve efficiency.

Furthermore, since the LLM output was referred to as a separate file in this demonstration experiment, there were comments that it would be better if it were included in the screen of the analysis system (in this experiment, Ghidra) or seamlessly linked to it.
Some participants also stated that, although the experiment was limited to static analysis logs, there were comments that the system could support other tasks, such as dynamic analysis logs.

Other issues raised included:

\begin{quote}
P2, P6
\textit{``Sometimes unimportant parts of the process are explained redundantly.''}

P2
\textit{``Prompt engineering is required to control the output of the LLM. This is a separate skill from the previous static analysis and must be learned anew.''}
\end{quote}

As described above, we have identified the challenges that need to be addressed when using an LLM output to support static analysis (RQ4).

\section{Discussion}
\label{sec:discussion}

\subsection{Feasibility of Using LLM as Static Analysis Assistant}

As demonstrated in Sections \ref{subsec:rq12} and \ref{subsec:rq3}, LLMs have the potential to be used as static analysis support, though they may not completely replace the existing analysis flow. 
Therefore, the skills and various supporting technologies for static analysis will likely continue to be required.

In addition, there were differences in the way the LLM was used depending on the experience of the analysts. 
P2 and P3, who had extensive experience with analysis, stated that they checked the LLM output after completing their own analysis. Specifically, they stated:

\begin{quote}
P2, P3
\textit{``I reviewed the explanation after analyzing the malware to see if there were any deviations from my understanding.''}
\end{quote}

In contrast, the remaining participants either checked the LLM output before analyzing the malware or analyzed the malware while also using the LLM output, though there were also individual differences. 
Although the current malware analysis flow is summarized for each tier level of the analyst in \cite{Wong2021}, one of our next steps is to derive a new analysis flow based on LLMs.

Further, as discussed in Section \ref{subsec:rq4}, confidentiality issues must be addressed. 
One possible method is to manually or automatically determine whether or not information is sensitive and send only non-sensitive information to an external LLM, but this is difficult to do. 
Regarding this method, P3, P4, and P5 said:

\begin{quote}
P3, P4, P5
\textit{``In particular, when sensitive information is encoded, it is not easy to judge, and there is a risk of mistakenly inputting sensitive information.''}
\end{quote}

The following comments were also made regarding sensitive information.

\begin{quote}
P6
\textit{``Sending sensitive information to ChatGPT is difficult due to the security rules of our organization.''}
\end{quote}

Therefore, a possible direction for future research is to build and improve the accuracy of local LLMs that can generate explanations without sending information externally. 
P2, P3, P4, P5, and P6 stated the following.

\begin{quote}
P2, P3, P4, P5
\textit{``A local LLM server would be useful.''}
\end{quote}

In addition, it has been pointed out that sentence generation using neural generative models such as GPT-4 used in this study causes hallucinations, in which the generated sentences differ from the facts \cite{Ziwei2023}.
This can lead to false positives that explain malware as if it has functions that it does not have, confusing the analyst and leading to incorrect conclusions.
For this reason, in addition to the accuracy of the LLM explanation, we also checked for the presence of hallucinations and false positives, but none were found in this experiment as far as the authors were able to confirm.
Retrieval Augmented Generation (RAG) is known as a method to effectively suppress hallucinations \cite{Towhidul2024}.
RAG is a method that retrieves documents related to input sentences from external data sources when LLM is queried and generates responses based on the related documents, thereby effectively suppressing hallucination.
In our experiments, as shown in Table \ref{tab:prompt_list}, explanatory text is generated based on the disassembled or decompiled results of the malware to be explained, and it is possible that hallucination is suppressed by the same effect as that of RAG.
However, the above is a hypothesis based on a single case study and should be verified quantitatively in the future with large data sets.

\subsection{Usability and Functionality}

As suggested by existing studies \cite{Mattei2022,mink2023}, usability is important, especially when introducing a new system into practice. 
During the interview, we asked about how the LLM was used during the analysis (or what kind of use is expected), the usability of the system to check the LLM output, and the functional requirements for this system. 
This section discusses usability and functionality, focusing on the content of these questions. 
Although we received 12 responses regarding usability and 21 responses regarding functionality, due to space limitations, we have selected some of the more important responses for discussion.
Note that all of responses regarding usability and functionality are listed on Table \ref{tab:codebook_interview} in Section \ref{sec:codebook_interview} of the appendix.

In this analysis, participants were provided with the system shown in Fig. \ref{fig:experimental_environment}, but because it was a separate screen from Ghidra, there were comments stating that it was cumbersome to go back and forth between the screens. Specifically, P3, P4, and P5 stated:

\begin{quote}
P3, P4
\textit{``Going back and forth between screens is cumbersome.''}

P4, P5
\textit{``When going back and forth between screens, it is difficult to tell what part of the code was being viewed.''}
\end{quote}

Therefore, in terms of usability, it is important to have the code on the same screen as the analysis system through plug-ins, or to be able to go back and forth smoothly, as is the case with similar existing tools. 
P2 mentioned the following for desirable features:

\begin{quote}
P2
\textit{``I would like a feature that reflects changes in analysis tools (renaming functions and adding comments) on the LLM side.''}
\end{quote}

This also suggests the importance of linking the LLM to existing analysis tools.
The following opinions were stated regarding the desired screens:

\begin{quote}
\textit{``Display of summary and suspicious areas on the homepage (e.g., dashboard).''}

\textit{``Summary display of malwares.''}

\textit{``Display functions in the graph view with explanations next to them.''}
\end{quote}

Since the explanations in this LLM are separated for each function, it is necessary for the analyst to determine the important functions after moving to each page one by one. 
In addition, since the explanation in this LLM was generated for each function, there were also requests to improve the clarity of the relationship with other functions.

In addition, although we provided static descriptions generated at specific prompts in this case, some participants used ChatGPT separately to query function descriptions for other prompts (P2) or to query the type of malware and possible related attack groups (P5). 
Therefore, it may be important to have an interface that can dynamically query LLM.

In this study, LLM explanations were provided in English and translated into Japanese, and all participants stated that they mainly used Japanese but also referred to English when necessary. 
P1 and P5 explicitly stated that it would have been better to have Japanese. 
Therefore, providing a translation function is also important for usability.

\subsection{Limitations and Future Work}

The generated text and various scores discussed in this paper are based on the model as of May 2023. 
As with existing similar studies, these are potentially subject to change or suspension in the future.

The malware used in this experiment was not completely new due to the need to prepare a solution. 
In addition, only one malware was used in the evaluation for simplicity. 
Therefore, there is a possibility that we were not able to evaluate the possibility of support for a completely unknown malware, and as well as the possibility of bias in the malware. 
We determined from the questionnaire that the malware was not known to the participants, and we eliminated the first possibility by prohibiting references to the analysis article on the web. 
In addition, the participants were instructed to respond to the questionnaire without limiting their answers to this malware, in order to suppress the second possibility.

As a first step in verifying whether the text generated by the LLM is practical, we evaluated the accuracy of the text itself (fluency, relevance, informativeness, and practicality). 
Although out of scope because of the simplicity of the system, we were able to obtain information from the analysts on functions that would be desirable to incorporate into the system for practical use. 
In the future, we would like to construct a system that incorporates these functions and evaluate the usability of the system using indicators such as the System Usability Scale \cite{SUS}.

In this study, several analysts commented that they were able to analyze the data more efficiently than without LLM explanations, suggesting that LLM explanations have effectiveness in supporting static analysis.
However, due to the focus of this study on the effects of LLM, all analysts had access to the LLM explanation, and no quantitative comparison was made with the analysis without the LLM explanation.
Since self-reported scores are known to have limitations, a quantitative comparison between cases where LLM explanations are accessible and those where they are not is a future challenge.
We believe that the results from the questionnaires and interviews in this study will provide valuable input for designing future quantitative evaluations.

The primary goal of this research is to reduce the extensive human intervention and high implementation of static malware analysis as described in Section 2.
In the questionnaire and interview, several participants stated that LLM seemed to reduce analysis time and analysis difficulty.
We believe this indicates that LLM could contribute to the above goals.
However, the evaluation in this paper does not directly assess the degree of human intervention and the time required for static malware analysis.
Therefore, future work should include direct measurement of analysis time and evaluation of the level of expertise by asking analysts of different levels to collaborate.

\subsection{Research Ethics}
The purpose and content of the study were explained to the participants in advance, and their consent to participate was obtained before the experiment was conducted. 
Pseudonyms have been used in this paper to prevent identification of individual participants. 
The handling and protection of personal information has been reviewed and approved by the privacy department of the author's organization. 
In addition, a Privacy Impact Assessment (PIA) was conducted under the supervision of our privacy department and confirmed that there were no issues.

\section{Related Work}

As mentioned above, while static malware analysis is important, its operation cost is high; thus, research is being conducted to support such analysis. 
For example, there are methods for assigning types and variable names to decompiled results based on function content \cite{Chen2022}. 
Studies have also aimed to clarify the mental model of reverse engineering through interviews with experimental participants and observations of analysis tasks \cite{Yakdan2016,Mantovani2023}. 
The method described in this paper further demonstrates the possibility of supporting static analysis from a different perspective by explaining functions using LLMs. 
It is also possible to use our method in combination with these existing methods to make the analysis more efficient.

LLMs are also being applied to various domains such as cybersecurity, for example, to fix vulnerable code \cite{Pearce2023} and find the root cause of incidents in cloud environments \cite{Ahmad2023}. 
Several LLM-based tools have also been developed, including Code Insight \cite{code_insight}, a service provided by VirusTotal that explains the functionality of malware written in PowerShell, as well as tools for pen-testing \cite{PentestGPT} and vulnerability scanning \cite{burpGPT}. 
Furthermore, there are research efforts to improve the readability and simplicity of decompilation using LLMs. 
The support methods described in this paper are inspired by these existing studies and tools. In addition, through collaborating with static analysts, this paper demonstrates the usefulness of LLMs.

Mink et al.'s study examined machine learning-based security tools and security practitioners' perceptions of these tools \cite{mink2023}. 
They found that low usability and a high number of false positives are disincentives for the adoption of machine learning-based tools in practice. 
These are consistent with the challenges identified by RQ4 of this paper, indicating the importance of overcoming these challenges for the introduction of LLMs into static analysis practice.

\section{Conclusion}

In this paper, we explore the possibility of using LLMs to support static malware analysis. 
Specifically, we evaluated the accuracy of LLMs by comparing analysis articles and LLM output, and determined that LLMs can provide practical accuracy. 
In addition, we performed a user study to verify the possibility of using LLMs as support and to identify areas for improvement for practical use in the future.
Our future work will include interviewing more static analysts and implementing a static analysis support system using an LLM.

\section{Acknowledgment}
The authors would like to thank the static analysts who participated in the user study to further this research.

This is a pre-print of an article published in Workshop on Security and Artificial Intelligence (SECAI 2024).

\bibliography{bib}
\bibliographystyle{splncs04}

\appendix

\section{Functions of Evaluation Malware}
\label{sec:appendix_a}

Table \ref{tab:malware_functions} shows the functions included in the malware used in the evaluation. 
For each function, we also indicate for each PID whether or not it was covered by the description in the LLM, for both decompilation and disassembly.

As described in Section \ref{subsec:eval_accuracy}, the functions of the malware and whether the LLM description covered each function were formulated by the authors through the coding process.

\section{Codebook of Interviews}
\label{sec:codebook_interview}
  
Table \ref{tab:codebook_interview} shows the codebook of the interview results.
The interviews consisted of seven items, as shown in Section 3.4.2.
Of these, except for two reminders (\textit{Reminder of the purpose of this interview.} and \textit{Reminder that the interview is being recorded.}) to the interviewees, all five items (\textit{Confirmation and detailed review of interview content (e.g., reasons for each answer).}, \textit{Challenges of using LLMs.}, \textit{How LLMs will be used in this analysis.}, \textit{Usability of the system for viewing LLM results and functional requirements of the system.}, and \textit{Any other topics.}) are listed in the codebook that was created.

We believe that the results presented in Table \ref{tab:codebook_interview} not only show the current state of LLM support for static malware analysis, but also lead to suggestions for future research topics.

\begin{table}[t]
  \begin{center}
  \caption{Features included in the malware for evaluation (babuk) and whether they are explained by LLM.}
  \scalebox{0.65}{
    \begin{tabular}{rlccccccccccc}
      \hline
      \hline
      \multirow{3}{*}{\#} & \multirow{3}{*}{Function} & \multicolumn{11}{c}{Explained by LLM?} \\
      \cline{3-13}
      & & \multicolumn{5}{c}{Decompile} & & \multicolumn{5}{c}{Disassemble} \\
      \cline{3-7}
      \cline{9-13}
      & & PID1 & PID2 & PID3 & PID4 & PID5 & & PID1 & PID2 & PID3 & PID4 & PID5\\
      \hline
      1 & Create a ransom note. & \checkmark & \checkmark & \checkmark & \checkmark & \checkmark & & - & \checkmark & - & \checkmark & -\\
      2 & Create a mutex. & - & \checkmark & \checkmark & \checkmark & \checkmark & & - & - & - & - & -\\
      3 & Provide a victim with a secret key generated as a random number. & - & - & - & - & - & & - & - & - & - & -\\
      4 & Encrypt a file. & \checkmark & \checkmark & \checkmark & \checkmark & \checkmark & & - & - & - & \checkmark & -\\
      5 & Assign an extension to the encrypted file. & \checkmark & \checkmark & \checkmark & \checkmark & \checkmark & & - & - & - & - & -\\
      6 & Erase the contents of the recycle bin. & \checkmark & \checkmark & \checkmark & \checkmark & \checkmark & & - & - & - & - & -\\
      7 & Delete a volume shadow copy. & \checkmark & \checkmark & \checkmark & \checkmark & \checkmark & & \checkmark & \checkmark & \checkmark & \checkmark & \checkmark\\
      8 & Enumerate logical drives. & \checkmark & \checkmark & \checkmark & \checkmark & \checkmark & & - & - & - & - & -\\
      9 & Enumerate network shares. & \checkmark & \checkmark & \checkmark & \checkmark & \checkmark & & - & - & - & - & -\\
      10 & Mount the identified drive. & \checkmark & \checkmark & \checkmark & \checkmark & \checkmark & & - & - & - & - & -\\
      11 & Exclude certain files/folders from encryption. & \checkmark & \checkmark & - & \checkmark & \checkmark & & - & - & - & - & -\\
      \hline
    \end{tabular}
  }
  \label{tab:malware_functions}
  \end{center}
\end{table}

\begin{table}[t]
  \begin{center}
  \caption{Codebook of interviews.}
  \scalebox{0.35}{
    \begin{tabular}{llll}
      \hline
      \hline
      Catergory & Subcatergory & Subsubcatergory & Codebook of Interviews\\
      \hline
      Confirmation and detailed review & About analysis & Analysis time & 5min, 60min, 80min, 90min, 120min\\
      \cline{3-4}
      of interview content.  &  & Analysis tools other than Ghidra that are regularly utilized & IDA Free\\
       &  &  & IDA Pro\\
       &  &  & ImmunityDebugger\\
       &  &  & x64dbg\\
       &  &  & LordPE\\
       &  &  & PE Bear\\
       &  &  & Stirling\\
       &  &  & CyberCheff\\
       &  &  & Wireshark\\
       &  &  & VirtualBox\\
       &  &  & Hopper\\
       &  &  & cuckoo sandbox\\
       &  &  & VirusTotal\\
       &  &  & Binary editor\\
       &  &  & Select the most appropriate tool for each malware.\\
      \cline{2-4}
       & LLM outputs & Fluency & 1--4\\
      \cline{3-4}
       &  & Reasons for feeling fluent & It's not weird at all.\\
       &  &  & It was so easy to read that just reading the output was enough to understand the process.\\
      \cline{3-4}
       &  & Reasons for not feeling fluent & There are parts where it is a question even though it is a comment.\\
       &  &  & Some sentences contained unnecessary information that made me feel weird.\\
      \cline{3-4}
       &  & Relevance & 1--4\\
      \cline{3-4}
       &  & Reasons for feeling relevant & They converted the parameters of WindowsAPI calls in an unambiguous way (e.g., 0x80 $\to$ FileHidden).\\
       &  &  & I don't think anything irrelevant was written.\\
       &  &  & The output was consistent with what the function did.\\
      \cline{3-4}
       &  & Reasons for not feeling relevant & Non-essential information is included redundantly.\\
       &  &  & Some content was too abstract.\\
       &  &  & Requires you to read the code yourself because of guesses, some of which are wrong.\\
       &  &  & Confusing explanations of unrelated items that are forcibly linked to malware behavior.\\
      \cline{3-4}
       &  & Informativeness & 1--4\\
      \cline{3-4}
       &  & Reasons for feeling informative & Maybe not perfect, but it was enough to assist.\\
       &  &  & Not too much information, not too little.\\
      \cline{3-4}
       &  & Reasons for not feeling informative & Information was closed to functions and missing caller/destination information.\\
       &  &  & Some long functions were too abbreviated.\\
       &  &  & Some parts did not directly answer the question.\\
       &  &  & Some parts were covered in outline but not in detail.\\
       &  &  & Some functions were explained line by line, but not in units of processes consisting of several lines.\\
      \cline{3-4}
       &  & Practicality & 1--4\\
      \cline{3-4}
       &  & Reasons for feeling practical & The description is useful.\\
       &  &  & It would be nice to have an overview of the function\\
       &  &  & It is fine as a function description\\
      \cline{3-4}
       &  & Reasons for not feeling practical & Difficult to rely only on LLM output for analysis.\\
       &  &  & Output is fine, but difficult to check caller/destination due to I/F.\\
       &  &  & Some of the details were wrong, while the overview was correct.\\
       &  &  & No overall flow or overview, so we had to figure it out ourselves.\\
       &  &  & If there is an obfuscation or other anti-analysis feature, it may not be well explained.\\
      \hline
      Challenges of using LLMs. &  & Challenges with sending information & No problem if only those that can be determined not to contain sensitive information are submitted\\
      \cline{3-4}
       &  & outside the organization via ChatGPT  & No problem if the samples are publicly available, such as in VT.\\
       &  &  & There is a possibility that information hard-coded into the sample (e.g., organization name, DNS, and authentication information) may be leaked.\\
       &  &  & If memory dumps are submitted, there is a possibility that confidential information may be included, resulting in a leak.\\
       &  &  & There may be contractual issues, such as not using external services.\\
      \cline{3-4}
       &  & Challenges due to ChatGPT's query count limit & May be difficult to submit a large number of samples at once, e.g. with a large number of features.\\
      \cline{3-4}
       &  & Challenges due to the fact that ChatGPT is fee-based & None (would submit if useful).\\
       &  &  & None (not a barrier).\\
      \cline{3-4}
       &  & Other challenges & Obfuscation will reduce accuracy.\\
       &  &  & Junk code will reduce accuracy.\\
       &  &  & Redundant explanation of unimportant parts (especially hard to read for beginners).\\
       &  &  & Considering the possibility that the LLM output is incorrect, it is necessary to secure it as well.\\
       &  &  & Difficult to use Japanese as training data when trying to do additional training.\\
       &  &  & Small screen.\\
       &  &  & Due to the character limit, the entire sample cannot be submitted, and the entire description cannot be generated easily.\\
       &  &  & Inefficient to make each query by yourself.\\
       &  &  & Need to control prompts (requires additional skills).\\
       &  &  & Input is interrupted, so accuracy may be compromised.\\
       &  &  & Many of the descriptions of prompts as malware were biased and suspicious.\\
       &  &  & Perhaps a more summarized and granular explanation is needed for beginners.\\
      \hline
      How LLMs will be used in this analysis. &  & How to use LLM in this user study & I used it to help me understand the function of a function.\\
       &  &  & I used it to determine if a function seemed important by skimming the function description.\\
       &  &  & I used it to get a preliminary understanding of the function before reading the decompiled code.\\
       &  &  & Identified a suspicious function from information about strings and imports, etc., and read the description of the function in question.\\
       &  &  & I extracted keywords from the LLM description and looked for important functions in the decompiler results.\\
       &  &  & After analyzing the sample, I checked the description to see if there were any deviations from my understanding.\\
       &  &  & After searching for main, read the function description by following the process with CALL, etc.\\
      \cline{3-4}
       &  & Other ChatGPT usage & Not used.\\
       &  &  & Queried sample type based on ransom note characteristics (starting with RAGROUP).\\
       &  &  & Retrieved the function name code.\\
       &  &  & Asked for the overall summary description of the sample,\\
       &  &  & Performed its own verification of the desired prompts,\\
      \cline{3-4}
       &  & Which decompiler/disassembler was used? & Decompiler only\\
       &  &  & Disassembler only\\
       &  &  & Both\\
      \cline{3-4}
       &  & Which did you see Japanese or English? & Japanese only\\
       &  &  & English only\\
       &  &  & Both\\
      \hline
      Usability of the system for &  & Preferred screen configuration & Place code and explanation on the same screen.\\
      viewing LLM results and functional &  &  & I want to see the function in a graph with an explanation next to it.\\
      requirements of the system. &  &  & Dashboard to show overview and suspicious areas.\\
       &  &  & Implement as a plugin for analysis tools and display on the same screen.\\
       &  &  & Display a summary for each function on one screen.\\
      \cline{3-4}
       &  & Features that would be nice to implement & Summary display of samples.\\
       &  &  & Comments function considering the caller/caller without closing to a function.\\
       &  &  & Display of important or suspicious parts.\\
       &  &  & Additional learning functions (insertion of analyst's findings and correction of incorrect parts).\\
       &  &  & Estimation of malware type (family name or sample type): VT can be wrong and static, making it difficult to get reasons and additional information.\\
       &  &  & Determine if the sample is a distributed or targeted type (to determine a course of action for further investigation).\\
       &  &  & Link to Windows API documentation.\\
       &  &  & Search function.\\
       &  &  & Ability to interactively ask questions.\\
       &  &  & Ability to suggest related families or attack groups based on the Windows API.\\
       &  &  & Ability to reflect changes in analysis tools (rename functions and add comments).\\
       &  &  & Ability to name functions.\\
       &  &  & Ability to estimate/explain what type of encryption should be used\\
       &  &  & Extraction of entry points and main body.\\
       &  &  & Estimate structure.\\
       &  &  & Correlation with dynamic and surface analysis results.\\
      \hline
      Any other topics. &  &  & Better than nothing, because it was used as a supplement.\\
       &  &  & Local LLM server would be useful.\\
       &  &  & Disassembly lacks competition.\\
       &  &  & When the screen goes back and forth, it is tedious to go back and forth, and it is difficult to know what part of the code you were looking at.\\
       &  &  & Decompiled code is harder to read than regular C, so an aid would be useful, especially for those who are not very good at parsing.\\
       &  &  & There was no more recent information (emerging ransomware and groups) than the information in ChatGPT ($\sim$2021).\\
       &  &  & It would be nice to be able to do fuzzy searches for devices etc. related to alerts/incidents.\\
       &  &  & Japanese language would be better.\\
       &  &  & Explanation of Windows API arguments (flags) was incorrect.\\
       &  &  & Where is this feature? It would be nice to be able to ask "Where is this function?\\
       &  &  & Explanation of \textit{GetWOW} process: This is a redundant explanation, although it is not important.\\
       &  &  & It is useful to summarize and display the results of string operations that should be tracked and reconstructed one by one.\\
       &  &  & The quality of the output depends on the performance of the decompiler (IDA Pro was able to decompile the samples more cleanly in this case).\\
       &  &  & Sending confidential information to ChatGPT is difficult due to the security rules of the organization.\\
       &  &  & It would be nice to have a service like ChatGPT with a high level of confidentiality.\\
       &  &  & It would be nice to have a general starting point for analysis (entry points, suspicious functions, etc.).\\
      \hline
    \end{tabular}
  }
  \label{tab:codebook_interview}
  \end{center}
\end{table}

\end{document}